# Superconducting Diode Effect in Gradiently Strained $Nb_{0.5}Ti_{0.5}N$ Films

Zherui Yang[1†]; Shengyao Li[1†]; Shaoqin Peng[2]; Xueyan Wang[1]; Liang Wu[3]; Ri He[2]; Zhen Wang[4]; Yanwei Cao[2*]; Xiao Renshaw Wang[1, 5*]

[1]Division of Physics and Applied Physics, School of Physical and Mathematical Sciences, Nanyang Technological University, Singapore 637371, Singapore
[2]Ningbo Institute of Materials Technology and Engineering, Chinese Academy of Sciences, Ningbo 315201, China
[3]Faculty of Material Science and Engineering, Kunming University of Science and Technology, Kunming, Yunnan 650093, China
[4]Department of Physics, University of Science and Technology of China, Hefei, Anhui 230026, China
[5]School of Electrical and Electronic Engineering, Nanyang Technological University, Singapore 639798, Singapore

[†] These authors contributed equally
* Emails: ywcao@nimte.ac.cn; renshaw@ntu.edu.sg

## Abstract

The superconducting diode effect (SDE), combining superconductivity with diode-like nonreciprocal current flow, recently emerges as an ideal candidate for zero-dissipation electronic circuits. Such technologically advantageous diodes are achieved by intricate material engineering to disrupt inversion symmetry, which leads to the production challenges as well as a limited pool of viable materials. Here we exploit the gradient interfacial strain to experimentally induce the SDE in $Nb_{0.5}Ti_{0.5}N$ (NTN) films grown on MgO substrates. Additionally, the SDE is tunable with an in-plane magnetic field and can be further enhanced by introducing an interfacial anisotropic pinning potential. Our findings establish interfacial strain gradient as a versatile tool for creating and enhancing tunable SDE.

**Significance Statement**

SDE promises an advancement toward zero-dissipation electronic circuits, yet their practical realization has been limited by the need for complex material engineering to break inversion symmetry. Our study introduces a novel approach by harnessing a gradient interfacial strain to induce the SDE in $Nb_{0.5}Ti_{0.5}N$ films on MgO substrates. We identified anisotropic pinning introduced by gradient interfacial strain as a previously unknown factor to enhance SDE. Our work paves the way for next-generation energy-efficient electronics and quantum devices, bridging fundamental advances in superconductivity with tangible technological innovation.

**Introduction**

The diode effect, fundamental phenomenon in solid-state physics enabling unidirectional current flow, underpins essential applications in modern electronics, such as current rectification, light emission, and protective circuits.[1-3] Recently, it has been discovered that diode effects can be enhanced during superconducting transitions, thereby garnering significant interest in inducing the SDE.[4-7] Due to its theoretically infinite on-off ratio, the SDE is anticipated to be crucial in the development of next-generation electronic devices.[8] Since its observation in 2020, SDE has been extensively explored in various superconducting systems, both experimentally and theoretically.[9, 10] A key requirement of the SDE system is the simultaneous breaking of inversion symmetry ($P$) and time-reversal symmetry ($T$), with the latter often achieved through an externally applied magnetic field or internal magnetism.[11-13] In some van der Waals materials, inherent lattice asymmetry allows for an intrinsic SDE.[14-16] Alternatively, inversion symmetry can be manually broken through specific stacking of material layers to disrupt the inversion symmetry, such as the superlattice superconducting films with alternating layers of niobium, vanadium, and tantalum.[17-20] However, van der Waals materials and stack engineering approach are greatly limited by the challenges of small fabrication scale and complexity in stacking engineering to break inversion symmetry, respectively.

Strain has been reported to be an effective method for tuning electronic properties of the material.[21-23] While applying strain alone preserves the material's inversion symmetry, introducing a strain gradient can induce polarization along its direction, breaking the inversion symmetry. This phenomenon has been widely utilized in applications such as enhancing the photovoltaic effect and inducing ferroelectricity.[24-26] One of the most potent approaches to introduce strain gradient is by growing epitaxial films thicker than their critical thickness, where interfacial strain is relaxed through the formation of misfit dislocations, creating a gradient. At the interface, strain is naturally introduced to accommodate the structural mismatch between the film and the substrate. This strain gradually dissipates as the lattice parameter of the film returns to its bulk value. Thus, selecting an appropriate substrate can break the inversion symmetry of the epitaxial superconductor film without compromising its superconducting performance.[27,28] Employing strain gradients in this way offers a promising avenue for realizing the SDE in superconducting films without the need for additional complex interventions.

In this study, we prepared NTN epitaxial films on single-crystalline MgO (001) substrates.[28] Unlike polycrystalline NTN films grown on Si (001) substrates, which preserve inversion symmetry, the NTN/MgO device exhibits strain gradients due to lattice mismatch at the NTN/MgO interface. Consequently, the SDE is observed in the NTN/MgO device, with its magnitude tunable by an in-plane (IP) magnetic field. Observations from second harmonic resistance suggest that the SDE is enhanced by an anisotropic pinning potential carried by dislocations near to the interface. The relation between the SDE and the anisotropic pinning potential revealed is, to our knowledge, has not been previously identified.

**Results and discussion**

The lattice parameter of the rock salt-type $Nb_xTi_{1-x}N$ compound ranges from $a$ = 4.244 Å (x=0) to 4.389 Å (x=1). Based on Vegard's law, the lattice parameter of $Nb_{0.5}Ti_{0.5}N$ is given as $a$ = 4.317 Å.[29] Similar to NTN, MgO also features a rock salt-type crystal structure with $a$ = 4.212 Å. As a result, there is a lattice mismatch of approximately -2.47% between NTN and the MgO substrate.[29,30] This mismatch induces a strain gradient perpendicular to the interface, breaking the inversion symmetry of the pristine NTN, which leads to structural polarization and facilitates the potential for a finite SDE magnitude. Fig. 1(a) schematically illustrates the SDE in the strained NTN film grown on MgO (001). The electrical current flows along the x-axis, while an IP magnetic field is applied along the y-axis. Under a magnetic field, both time reversal and inversion symmetry are broken. As a result, the charge carriers are Cooper pairs in one direction but electrons in the opposite direction, leading to the realization of SDE.

Fig. 1(b) presents the temperature-dependent resistance of the strained NTN. The thickness of the sample is around 100 nm estimated by the deposition rate and time. The width of the channels is 5 μm. The IP and OOP coherence lengths $\xi_{ab}$ and $\xi_c$ of the film is estimated as $\xi_{ab}$=15 nm and $\xi_c$=19 nm respectively, at $T$ = 11.5 K (Fig. S8). The penetration depth $\lambda$ depends on the Ginzburg Landau parameter κ.[31] As a strong type II superconductor, κ is estimated to be 30 in NTN film, which leads to $\lambda$ around 400 nm based on the relation $\lambda= \kappa\xi$.[32] The sample exhibits metallic behavior until a superconducting transition at a critical temperature ($T_c$) of 14.9 K, defined at the midpoint of the transition. Fig. 1(c) shows the asymmetric *V-I* relationship observed during current sweeps in opposite directions. current sweep in opposite directions. For a forward sweep (red curve), the current follows the sequence: zero to positive (0 - max), positive to negative (max - min), and negative back to zero (min - 0). Conversely, the negative sweep (blue curve) follows the opposite sequence. Despite these directional differences, the two curves overlap, and both display asymmetry with respect to the *I* = 0 axis. This observation confirms that the asymmetric *V-I* characteristics is an intrinsic property of the strained NTN film and not a result of Joule heating.

To illustrate the influence of external magnetic fields on critical current, Fig. 1(d) presents the *I-V* curves obtained by sweeping the current in opposite directions under IP magnetic fields ranging from 0 T to 0.2 T. Under zero magnetic field (black curves), the absolute value of the critical current under the positive current sweep, $|I_{c+}|$, is equal to its counterpart under the negative current sweep, $|I_{c-}|$. Remarkably, the nonreciprocity of the critical current is controllable under an IP magnetic field. Despite the simultaneous suppression of $|I_{c+}|$ and $|I_{c-}|$ under an external magnetic field, $|I_{c+}|$ is always higher than $|I_{c-}|$. This magnetic field tunable nonreciprocity promises the functionality of the SDE, where the device switches between the resistive and superconducting states by altering the current and magnetic field directions. Fig. 1(e) exhibits the phenomenon by applying a sequence of square wave current pulses, at an amplitude of ±800 µA, a frequency of 0.05 Hz (upper panel). The lower panel records the corresponding changes in longitudinal voltage. Under a magnetic field along x-axis, a resistive state is induced for positive current bias, while a superconducting state persists under negative current bias (red line). Conversely, reversing the direction of the magnetic field results in the opposite switching behavior (blue line), consistent with previous findings.[10] The magnetic controllable switching between superconducting and resistive phases is the so-called superconducting diode effect. Compared to junction-based SDE, the strained NTN film is advantageous in durability. Fig. 1(f) presents the repetitive modulation for over 500 cycles, using a 0.17 Hz square wave pulse train (upper panel). The device reliably alternates between superconducting and resistive states without any detectable degradation, demonstrating the robustness of the SDE.

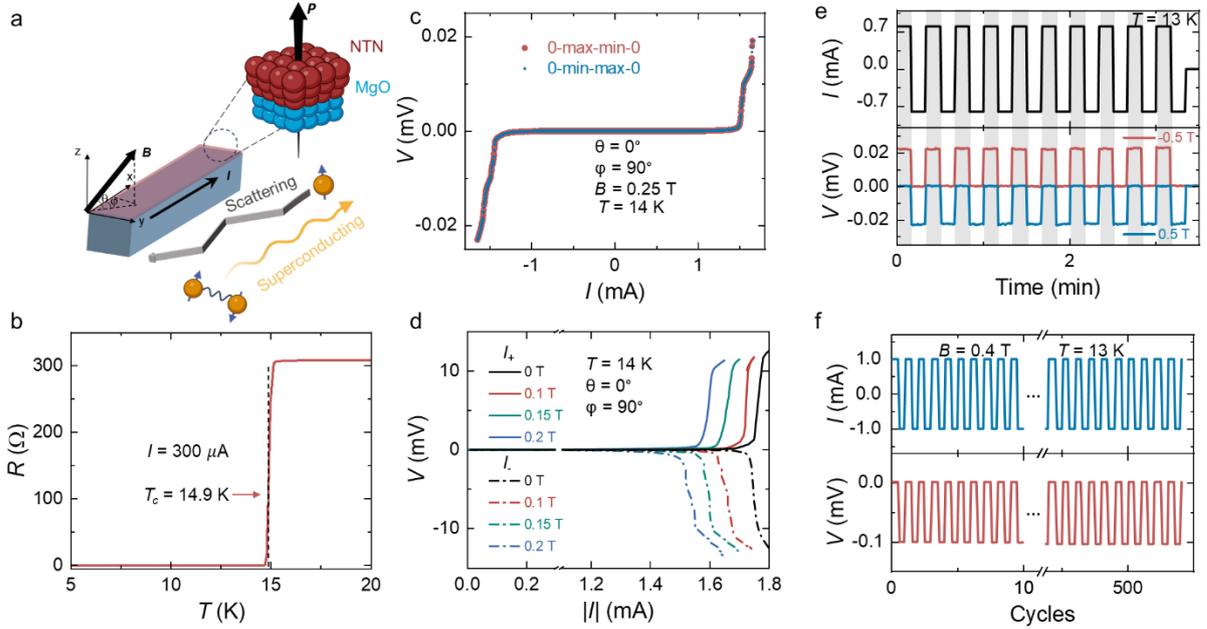

Fig.1 Illustrations of the SDE device. a) Illustration of SDE within the NTN / MgO heterostructure. θ and φ are polar and azimuthal angles respectively. b) Temperature-dependent resistance around superconducting transition. The critical temperature ($T_c$) is 14.9 K, defined as half of the superconducting transition. c) *V-I* relationship under positive(red) and negative(blue) current sweep under an IP magnetic field *B* = 0.25 T at *T* = 14 K. d) *V-I* characteristics of positive and negative current sweep directions under different magnetic fields at *T* = 14 K. e) Alternative switching between resistive state and superconducting state under a train of current pulses at *B* = 0.5 T, *T* = 13 K. f) Long-term cyclic reliability of the SDE.

We further explore the magnetic-field dependence of the SDE in this strained NTN film. Fig. 2(a) presents the magnetic field dependence of the critical currents under IP magnetic fields at *T* = 14 K. The application of an external magnetic field tends to break the superconductivity, with the suppression of $I_c$ for the current of both directions. Specifically, both $I_{c+}$ (blue curve) and $I_{c-}$ (red curve) decrease with increasing magnetic field. Notably, within the range of 0.5 T, the response of $I_c$ to the magnetic field exhibits directional dependence, with the magnitudes of $I_{c+}$ and $I_{c-}$ interchanging upon reversing the direction of the magnetic field. As the magnetic field continues to increase, $I_{c+}$ and $I_{c-}$ converge due to the overall suppression of superconductivity. The strength of the SDE is evaluated by the difference between $I_{c+}$ and $I_{c-}$, denoted as $\Delta I_c = I_{c+} - I_{c-}$. Fig. 2(b) shows $\Delta I_c$ as a function of the magnetic field. Here, $\Delta I_c$ exhibits an antisymmetric behaviour for the magnetic field, being negligible at $B$=0 T, peaking at ±0.26 T, and gradually diminishing as the superconductivity is further suppressed. For SDE that originated from polar systems, the effect is maximized when the magnetic field is orthogonal to the current and polarization.[33] To extract this directional behaviour, Fig. 2(c) presents the angle-dependent $\Delta I_c$ as the magnetic field rotates within the $xz$ plane. $\Delta I_c$ is negligible except near IP, where two distinct peaks with opposite signs emerge, indicating that the SDE in strained NTN is predominantly generated and controlled by the IP magnetic field. Fig. 2(d) charts the temperature dependence of $\Delta I_c$ around the $T_c$. As the temperature increases, $\Delta I_c$ first rises to a maximum of 220 μA at 13.7 K, before gradually diminishing to zero above $T_c$. These results indicate that the SDE in the sample is finely tunable through adjustments of magnetic field and temperature.

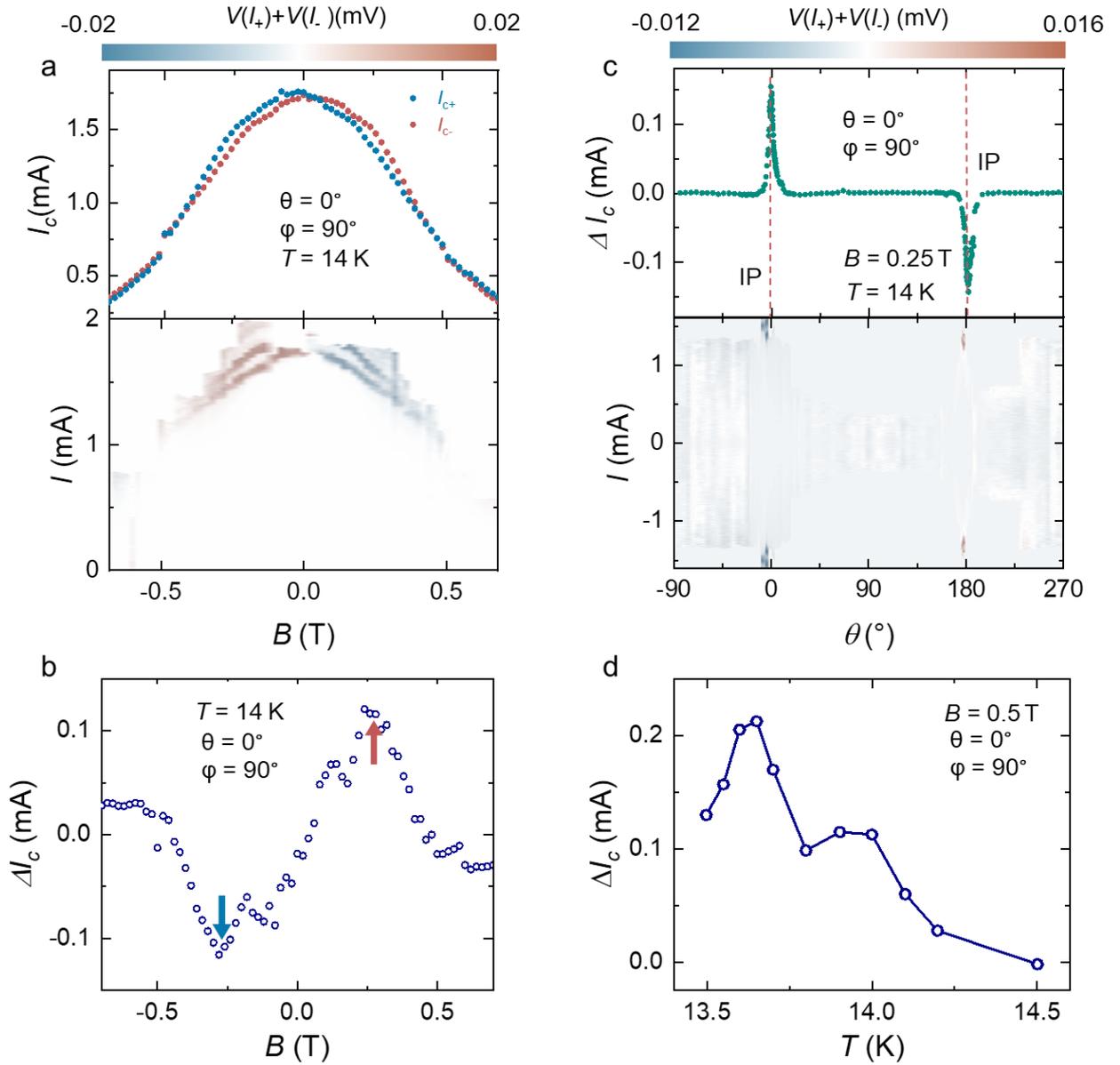

Fig. 2 Magnetic field tunable SDE in strained NTN films. a) Critical currents as a function of magnetic field, and the contour plot of the rectification voltage signal at 14 K. b) Magnet field dependence of $\Delta I_c$ at $T$ = 14 K. c) The angular dependence of $\Delta I_c$ and the contour plot of the rectification voltage signal at 14 K at $B$ = 0.25 T. d) Temperature dependence of $\Delta I_c$ under an IP magnetic field $B$ = 0.5 T.

The mechanism of the field tunable SDE in NTN film can be understood in the following manners. According to flexoelectric theory, in the presence of a nonuniform strain u, there is an electric field $E_{flexo}$ described by $E_{flexo} = f_{eff} \partial u/\partial x$, $f_{eff}$ is the flexo-coupling coefficient and $x$ is the direction of the strain gradient.[34, 35] This electric field, which is along the strain gradient direction, naturally introduces a Rashba type spin orbit coupling (SOC) term to the material.[36, 37] For small electric fields, we can assume that the electric field is proportional to the Rashba coefficient $α$, i.e., $α \propto E_{flexo}$.[37] Therefore, to illustrate the relationship between the strain gradient and SDE, it is sufficient to consider the interplay between Rashba SOC and SDE. To this end, we consider a Hamiltonian describing a s-wave superconductor with SOC:

$$H(\alpha) = \left(\psi_{k\uparrow}^{\dagger}, \psi_{k\downarrow}^{\dagger}\right)\left(\frac{k^2}{2m} - \mu + \alpha(\hat{z} \times \boldsymbol{k}) \cdot \boldsymbol{\sigma} - \boldsymbol{B} \cdot \boldsymbol{\sigma}\right)(\psi_{k\uparrow}, \psi_{k\downarrow})^T + \Delta\psi_{k\uparrow}^{\dagger}\psi_{-k\uparrow}^{\dagger} + \text{h.c.}, \quad (1)$$

where $\psi_{k\uparrow}^{\dagger}$ and $\psi_{k\uparrow}$ are creation and annihilation operators of electrons respectively, $k^2/2m - \mu$ is the kinetic term, $\boldsymbol{k} = (k_x, k_y)$ is the in-plane momentum, $m$ is the effective mass of electrons, $\mu$ is the chemical potential, $\boldsymbol{B}$ is the magnetic field, $\boldsymbol{\sigma} = (\sigma_x, \sigma_y, \sigma_z)$ are Pauli matrices represent for spin and $\Delta$ is the superconducting order parameter. The Rashba SOC combined with a magnetic field along y-axis results in asymmetric band structure with respect to $k_x$ (Fig. S15). With some symmetrical considerations, it is easy to show that the $\Delta I_c(\alpha) = -\Delta I_c(-\alpha)$, which means the direction of SDE is also determined by the Rashba SOC, and the magnitude of SDE is therefore related to the strain gradient by $\Delta I_c \propto \partial u/\partial x$.[38] In a word, strain gradient induces Rashba SOC, which controls the magnitude of SDE. The direction of SDE depends on the direction of strain gradient—whether the strain introduced by lattice mismatch is compressive or tensile. Fig. 3(a) provides a schematic illustration of spin splitting in the energy bands of both pristine NTN and strained NTN on MgO substrates. In this epitaxial NTN film, structural asymmetry arises due to the strain gradient caused by lattice constant mismatch between the NTN film and the MgO substrate. When the initial layer of NTN grows epitaxially on the MgO (001) surface, it adopts a lattice structure similar to that of the substrate, resulting in interfacial strain. As the film grows thicker, it gradually relaxes into its bulk configuration, and this strain relaxation creates a gradient that induces the flexoelectric field perpendicular to the interface, ultimately giving rise to Rashba spin-splitting in the energy bands. This gradient-induced breaking of inversion symmetry, when combined with time-reversal symmetry breaking from the applied magnetic field, results in the observed nonreciprocity of the critical current. Fig. 3(b) presents the reciprocal space mapping (RSM) to verify the presence of interfacial strain. The reciprocal space vectors $Q_{[001]}$ and $Q_{[100]}$ correspond to the OOP and the IP directions respectively. The separation between the profiles of NTN (lower left) and MgO (upper right) indicates a substantial lattice mismatch. The reciprocal space vectors $Q_{[001]}$ and $Q_{[100]}$ correspond to the OOP and the IP directions respectively. From the RSM, the lattice parameter of the NTN film (a = 4.323 Å) is close to its bulk value (a = 4.317 Å), suggesting the relaxation of NTN film with the increase in thickness.[39] The diffraction profile is homogeneous for MgO, while is elongated along the (201) direction for NTN, indicative of strain within the NTN film.[40] According to high-resolution XRD measurements, no secondary phase is formed despite the considerable interfacial strain between the film and the substrate (see Fig. S2).[41] To further investigate the strain gradient in the NTN/MgO device, we utilize a scanning transmission electron microscope (STEM) to extract both the IP and OOP components of strain tensor around the interface.[42] As shown in Fig. 3(c), near to the interface, NTN can be divided into two regions by the dashed line. The region between the interface and the dashed line is partially strained until fully relaxed at the dashed line. The IP strain mapping shows that the IP dislocations concentrate in this region. On the other hand, the OOP lattice parameters exhibit minimal variations and there is no observation of OOP dislocation (Fig. S5). The result shows that the strain gradient is along OOP and mainly exists within approximately 40 nm from the interface, which matches our transport measurement. Also, we observe the IP misfit dislocations in the gradiently strained region of NTN, which is due to the lattice mismatch will be discussed later.[44]

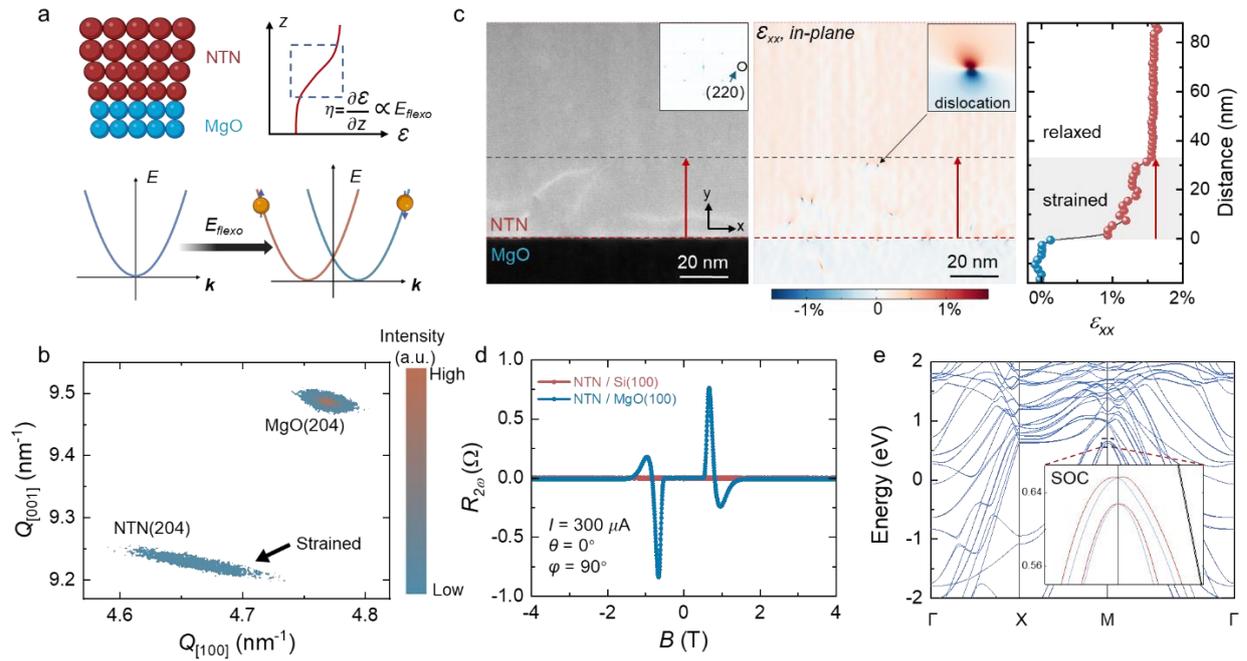

Fig. 3 Strain gradient in NTN/MgO device. a) Schematic illustration of the Rashba spin splitting induced by the strain gradient in the NTN film grown on the MgO substrate. b) Reciprocal space mapping (RSM) of NTN film on MgO substrate around (102) Bragg diffractions. c) STEM image of the NTN/MgO sample, the IP strain

components derived from (220) diffraction spots using Geometric phase analysis method, and strain profiles as a function of distance from the interface. The red arrows specify the inhomogeneous strain region near to the NTN/MgO interface. The Inset highlighting details of misfit dislocations at and near the interface. d) Second-harmonic magnetoresistance of NTN films on Si (red) and MgO (blue) substrates. e) Band structure of strained NTN layers. The inset zooms in the band structure near the M point.

The role of strain gradient on SDE is further verified by additional experiments conducted on epitaxial NTN films on MgO and Si substrates. In contrast to the strained crystalized NTN on MgO substrate, the interfacial strain of the polycrystalline NTN on Si can be ignored. Fig. 3(d) compares the second harmonic magnetoresistance ($R_{2\omega}$) between the two NTN films, which is proportional to the magnitude of magnetochiral anisotropy (MCA) and is therefore, consequently, directly related to SDE.[10,15] At 11.5 K, the $R_{2\omega}$ of the NTN/MgO reaches a maxima of 0.8 Ω under a 0.4 T IP magnetic field (blue curve). In contrast, no signal was detected in the NTN/Si system under identical conditions (red curve), indicating that SDE is exclusive to strained NTN films (Fig. S13).

The breaking of inversion symmetry in strained NTN films leads to spin splitting in their energy bands, which, when combined with the breaking of time-reversal symmetry, results in the SDE and nonreciprocal transport behavior.[44] To visualize the influence of interfacial strain on band structure, density functional theory (DFT) calculations were performed on both unstrained and strained NTN layers. Fig. 3(e) shows the band structure of the NTN/MgO heterostructure along a high-symmetry line *Γ-M-X-Γ*. The zoomed-in plot near the **M** point (Fig. 3(d) inset) highlights the spin-orbit coupling-induced band splitting (blue and red curves). In contrast, no band splitting is observed in unstrained NTN layers where the centrosymmetric lattice structure is preserved (Fig. S14). As demonstrated earlier, the strain gradient in our system breaks inversion symmetry along the out-of-plane direction. The DFT calculations further confirm that this structural asymmetry enables Rashba-type spin-splitting, which is the underlying mechanism for the observed SDE.[10]

Although distinct phenomena, both SDE and nonreciprocal transport originate from MCA.[17] Nonreciprocal transport detects the nonlinear resistance $R_{2\omega}$, which is related to the MCA coefficient γ by $R_{2\omega} \propto \gamma BI$, where *B* and *I* are magnetic field and current respectively.[1] Therefore, $R_{2\omega}$ provides valuable insights into the behavior of γ under varying magnetic fields and temperatures.[45] To this end, we conducted a.c. measurement on NTN and investigate both first and second harmonic magnetoresistances. This approach allows the separation of normal and nonreciprocal resistances. Details on the configurations of the a.c. measurement can be found in Method. Based on the above discussion, the inversion symmetry of NTN/MgO device is broken by the NTN/MgO interface, with the polarization oriented perpendicular to the interface. This can be verified by the angular dependence of $R_{2\omega}$. To this end, NTN films were etched into a sunbeam pattern (inset of Fig. 4(a)) to investigate the directional dependence of nonreciprocal transport relative to the magnetic field. Fig. 4(a) illustrates the angle-dependent $R_{2\omega}^{max}$ when the magnetic field rotates in *φ* plane, where $R_{2\omega}^{max}$ is the maximum value of $R_{2\omega}$ at each $\varphi$. $R_{2\omega}$ vanishes when the current is parallel to the magnetic field ($\varphi = 0°$) and reaches its maximum when $\varphi = 90°$, where $\varphi$ represents the angle between the magnetic field and the current direction. On the other hand, when the magnetic field rotating in $\theta$ plane, $R_{2\omega}$ reaches its maximum with a sharp peak when the field is parallel to the sample plane but vanishes when the field is oriented along the z-axis (Fig. S10). Therefore, the electrical voltage of NTN/MgO film follows the phenomenological equation

$$V = R_0 I(1 + \gamma(B \times z) \cdot I), \qquad (2)$$

where $R_0$ is the linear resistance of the material and $\gamma$ represents the MCA coefficient. These results suggest that the strained NTN film can be categorized as an archetype Rashba superconductor, with the polarization oriented at OOP.

Near the superconducting transition, changes in vortex motion can give rise to exotic temperature dependencies of $R_{2\omega}$.[46] Temperature-dependent $R_{2\omega}$ measurements were conducted to investigate the constituents of these nonreciprocal phenomena within the strained NTN film. Fig. 4(b) shows the $B-T$ diagram of $R_{2\omega}$ under an IP magnetic field, where the evolution of $R_{2\omega}$ with respect to the temperature is clearly exhibited. At low temperatures, $R_{2\omega}$ exhibits a single contribution, and its magnitude decreases monotonically as temperature increases until 11 K, where the red signal sharpens and an opposite signal emerges at higher magnetic fields. As the temperature increases further, $R_{2\omega}$ first reaches its maximum at 11.7 K, with $\gamma_{max}$=618 T$^{-1}$A$^{-1}$, and then declines until it vanishes at 12.1 K. Based on the above observations, we separate the diagram in following regions. Under a small magnetic field, vortices in the NTN film are bounded like a solid and the resistance is zero. With increasing magnetic field, at lower temperatures (<11 K), NTN transits from a vortex solid state to a vortex liquid state, where finite resistance arises from the flow of unbound vortices. In this region, $R_{2\omega}$ is solely attributed to the vortex motion and decreases as temperature rises, due to the weakening of the pinning sites. With further temperature increase, the vortex lattice melts, and vortex motion becomes thermally activated under small magnetic fields, a phenomenon known as thermally activated flux flow (TAFF).[47, 48] In this regime, we observe an enhancement of $R_{2\omega}$, corresponding to the sharp peak in Fig. 4(b). Close to the superconducting transition temperature $T_c$, superconducting fluctuations also contribute to nonreciprocal transport, inducing a phenomenon known as paraconductivity, which we attribute to the negative signal in Fig. 4(b). The combined effect leads to the multiple reversal of $R_{2\omega}$ with increased amplitude at higher temperatures.[45] To explore the mechanism of the enhanced $R_{2\omega}$ at TAFF region, the $\theta$ – $B$ phase diagram for $R_{\omega}$ and $R_{2\omega}$ are shown in Fig. 4(c) and 4(d), respectively. Although the angular dependence of the upper critical field (red dashed line) shows a rather smooth pattern, the lower critical field (black dashed line) exhibits a sudden enhancement near $\theta$ = 180°(IP). This sharp angular dependence around the IP direction indicates the presence of highly anisotropic pinning potentials.[49-51] The second harmonic signal attributed to TAFF (liquid-drop-like region) is located exactly between the upper and lower critical fields, suggesting that the enhancement of $R_{2\omega}$ at this region is related to the anisotropic pinning potential. As shown in Fig. 3(c), misfit dislocations were observed distributed around the partially strained region of the NTN film. In epitaxial films, these dislocations are generated to release the misfit strain induced by lattice mismatch and serve as pinning cores.[43] Notably, the angular and temperature dependencies of the SDE closely resemble those of the TAFF-induced second harmonic signal. These results indicate that the lattice mismatch not only induces SDE and nonreciprocal transport by breaking inversion symmetry, but also enhances these effects through the generation of anisotropic pinning potentials.[46]

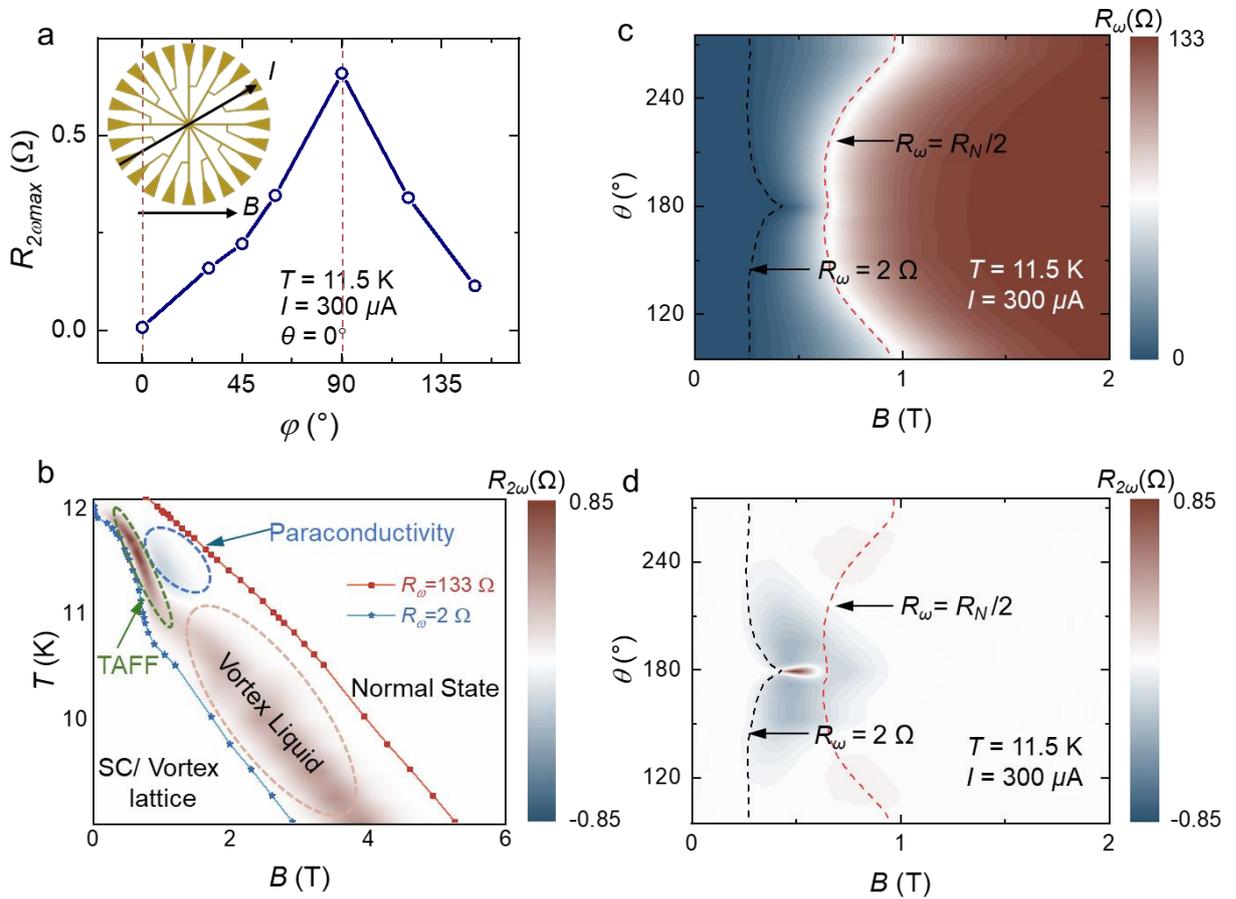

Fig. 4 Nonreciprocal charge transport of the NTN/MgO device. a) $R_{2\omega}^{max}$ as a function of the relative angle ($\varphi$) between the current and the IP magnetic field. b) Phase diagram of $R_{2\omega}$ in the $B$-$T$ plane under IP magnetic fields. The blue circle and red diamond lines represents for the breakdown of superconducting phase and recover to normal state $R_N$ = 133 Ω, respectively. The nonreciprocal transport behavior is detectable throughout the region. c) Superconducting phase diagram in the $B - \theta$ plane at 11 K under current $I$ = 300 μA. The black dashed line and red dashed line correspond to lower and upper critical field, respectively. d) Phase diagram of $R_{2\omega}$ in the $B - \theta$ plane at 11 K under current $I$ = 300 μA.

**Conclusions**

In conclusion, we demonstrate SDE in a gradiently strained NTN film. The nonreciprocity of critical current is well established, with the direction of the rectification effect reversing under opposite magnetic fields. The existence of the strain gradient and the induced OOP polarization is supported by IP strain component mapping and the angular dependence of SDE. These results indicate that the lattice mismatch in epitaxial films could indeed induce SDE. Additionally, misfit dislocations, observed near the interface, release interfacial strain and act as pinning cores for vortices, enhancing the SDE when the magnetic field is precisely IP. Unlike conventional SDE systems that rely on complex fabrication procedures, the SDE in strained film is readily achieved by utilizing the lattice mismatch between superconducting materials and substrates. This study establishes interfacial strain gradient as a novel and effective approach in realizing SDE, thus broadening the family of SDE and advancing the development of SDE devices.

**Supplementary Materials**

See the supplementary materials for sample characterizations (optical images, X-

ray diffraction, Raman spectroscopy, energy-dispersive X-ray spectroscopy) of NTN films, transport measurements demonstrating 2D superconductivity and nonreciprocal behavior, and theoretical discussions linking strain gradient to the SDE via Rashba spin-orbit coupling.

**Acknowledgements**

Z.R.Y. thanks Yue Mao for the valuable discussion. Z.W. acknowledges support from National Natural Science Foundation of China (No. 12304035). The authors thank the Instruments Center for Physical Science at USTC, Hefei, Anhui, China, for assistance with STEM experiments. X.R.W. acknowledges support from Singapore Ministry of Education under its Academic Research Fund (AcRF) Tier 1 (Grant No. RG82/23) and Tier 2 (Grant No. MOE-T2EP50220-0005). This research is also supported by the Singapore Ministry of Education (MOE) Academic Research Fund Tier 3 grant (MOE-MOET32023-0003) "Quantum Geometric Advantage". X.R.W. designed and directed this study.

**Methods**

**Material Growth**

Epitaxial $Nb_{0.5}Ti_{0.5}N$ films were synthesized on single crystalline (001)-oriented MgO substrates (5 × 5 × 0.5 mm$^3$) by a homemade high-pressure radio frequency (RF) magnetron sputtering according to the method in previous paper.[52] Briefly, the RF power was set at 100 W. The purities of 2-inch $Nb_{0.5}Ti_{0.5}$ targets and $N_2$ reactive gas are 99.995% and 99.999%, respectively. The base vacuum pressure before growth is ca. 3 × 10$^{-8}$ Torr. The $N_2$ pressure during growth was kept at 0.005 Torr with a gas flow of 1.7 sccm. The MgO substrates were mounted on a SiC absorber which was heated by a laser on the backside. The growth temperature was 1100 ℃ which was monitored by an infrared pyrometer on the backside SiC absorber. To guarantee the uniformity of films, the sample stage was rotated at a speed of 5 rpm when the film depositing. After growth, the films were cooled down to room temperature at a cooling rate of 25 ℃ per minute in the same atmosphere as deposition.

**Device Fabrication**

NTN film was patterned by an Ultraviolet Maskless Lithography machine (TuoTuo Technology (Suzhou) Co., Ltd.) and then etched with an Argon ion beam by the ATC-2020-IM ion milling system to take four probe transport measurements. The Argon ion beam removed the spare material and pattern to the hall bar and sunbeam shape for SDE and nonreciprocal transport measurements respectively. The width of the channel is around 5 μm.

**Electrical measurement**

A Keithley 2450 was used to apply a d.c. current to the device and the longitudinal d.c. voltage was measured by Keithley 2000 voltmeter. The $R_\omega$ and $R_{2\omega}$ were measured simultaneously using a Keithley 6221 source meter and two SR830 lock-in amplifiers with a 137.7 Hz sine wave AC current. Both the AC voltage and the current are measured in its effective (RMS). Based on the principle, we symmetrized the $R_\omega$ raw data and antisymmetrized the $R_{2\omega}$ raw data.

**Characterization of the sample**

The crystal structure and lattice parameters of the NTN film were characterized by a high-resolution X-ray diffraction (XRD) diffractometer (Bruker D8 Discovery) with the Cu Kα source (λ = 1.5405 Å).

The STEM and X-ray energy dispersive spectroscopy (EDS) experiments were conducted on a 300 kV Themis Z microscope equipped with a spherical aberration corrector for the condenser lens. The atomic resolved high-angle annular dark-field (HAADF) STEM images were collected with a 25 mrad convergent angle and a collection angle of 41–200 mrad.

**Details of band structure calculation**

The DFT calculations were performed using a plane-wave basis set with a cutoff energy of 550 eV as implemented in the Vienna Ab initio Simulation package (VASP),[53] and electron exchange-correlation potential was described using generalized gradient approximation and Perdew-Burke-Ernzerhof solid scheme.[54] The Brillouin zone was sampled with a 12×12×1 MonkhorstPack k-point grid. To consider the interface of MgO and TiNbN, a freestanding film of $(MgO)_2|(TiNbN)_4$ was constructed. The vacuum region spans 40 Å to prevent coupling between periodic images. The 1×1 lattice in the x-y plane was adopted. The optimized lattice constant of freestanding film is 4.21 Å. SOC is duly considered in electronic band structure calculations.